%Paper: hep-ph/9304268
%From: FTLIPKIN@WEIZMANN.weizmann.ac.il
%Date: 19 Apr 93   14:54 +03

%macropackage=phyzzx
\vsize=7.5in
\hsize=6.6in
\hfuzz=20pt%this is because one equation is a bit wide.
\tolerance 10000

\baselineskip 12pt plus 1pt minus 1pt
\pageno=0

\def\){]}
\def\({[}
\def\sno{\smallskip\noindent}

\def\rjustline#1{\line{\hss#1}}

\rjustline{WIS-93/34/Apr-PH}
\rjustline{TAUP XXXX-93}

\centerline{\bf
A QCD Relation Between Hyperfine Splittings in Isospin doublets}
\centerline{\bf A Light Quark Symmetry?}
\author{Harry J. Lipkin}
\sno
\centerline{Department of Nuclear Physics}
\centerline{\it Weizmann Institute of Science}
\centerline{Rehovot 76100, Israel}

\centerline{and}

\centerline{School of Physics and Astronomy}
\centerline{Raymond and Beverly Sackler Faculty of Exact Sciences
}
\centerline{\it Tel Aviv University}
\centerline{Tel Aviv, Israel}

\centerline{April 19, 1993}
\vskip 0.2in

\def \cn{Collaboration}
\def \ite{{\it et al.}}
\def \Sov#1#2#3#4#5{Zh. Eksp. Teor. Fiz. {#1}(#3)#2; Sov. Phys. JETP
{#4}(#3)#5.}

\def \plb#1#2#3{Phys. Lett. B {#1} (#3) #2.}

\def \prd#1#2#3{Phys. Rev. D {#1} (#3) #2.}
\def \prl#1#2#3{Phys. Rev. Lett. {#1} (#3) #2.}

\abstract

Recent data on the $D^*$ masses confirm the QCD prediction based on the
\REF{\ICHJL}{Isaac Cohen and Harry J. Lipkin,  \plb{84}{323}{1979}}
very general assumption $\({\ICHJL}\)$ that the spin-dependence of the quark
couplings to photons and gluons have the same structure and differ only in the
value of the coupling constants,
This prediction with no free parameters and using the
$\Xi$ and $\Xi^*$  masses as input follows from noting that
the charmed antiquark  $\bar c$ and the strange diquark (ss) have exactly the
same couplings to both the photon (Q=-2/3) and to the gluon (both are
color antitriplets). New predictions are obtained with this approach.
Corrections to the naive model are investigated and shown to be small.

\endpage

Recent very precise measurements of the isospin splittings in the charmed
pseudoscalar and vector meson multiplets
\REF{\CLEO}{ CLEO \cn, D. Bortoletto \ite, \prl{69}{2046}{1992}}
$\({\CLEO}\)$ combined with
other experimental masses\REF{\PDG}{ Particle Data Group, J. J.
Hern\'andez \ite, \plb{239}{1}{1990}}$\({\PDG}\)$
confirm a nearly
model-independent QCD prediction$\({\ICHJL}\)$,
$$ (10.3 \pm 0.6 \pm 0.4)\times 10^-3
= {{\delta m(D) - \delta m(D^*) }\over{M(D^*) - M(D)}}=
{{\delta m(\Xi) - \delta m(\Xi^*)}\over{M(\Xi^*) - M(\Xi)}}
= (15 \pm 4)\times 10^-3
\eqno(1) $$
where $\delta m(h)$ denotes the mass splitting in a given hadron isospin
doublet denoted by $h$. This prediction with no free parameters follows
from noting that the charmed antiquark  $\bar c$ and the strange diquark
(ss) have exactly the
same couplings to both the photon (Q=-2/3) and to the gluon (both are
color antitriplets). When a light u or d quark is added to a $\bar c$
to make a $D$ or $D^*$ or added to a (ss) pair to make a $\Xi$ or $\Xi^*$
the strong and electromagnetic contributions to the hyperfine splittings
have the same ratio in both systems.
The hyperfine interaction of both of these systems with an
additional light quark thus changes by the same ratio when the light
quark flavor is changed from $u$ to $d$.
This spin dependence of hadron isospin mass splittings
has been used by Sakharov\REF{\Sakh}{ A. D. Sakharov,\Sov {79} {350}
{1980} {52} {175}.}
$\({\Sakh}\)$
in the DGG model\REF{\DGG}{ A. De R\'ujula, H. Georgi, and S. L. Glashow,
\prd{12}{147}{1975}}
$\({\DGG}\)$
to obtain an estimate for the ratio $\alpha_{strong}/\alpha$.
It has also been used with new data\REF{\JLR}{James Amundson \ite,
\prd{47}{3058}{1993}}
$\({\JLR}\)$ to obtain meson decay constants.

The above derivation is more general than the previous
version$\({\ICHJL}\)$ which used additional assumptions to treat
other hadron masses. Our very general assumption
that the spin-dependence of the quark
couplings to photons and gluons have the same structure and differ only
in the value of the coupling constants can be applied simply to all color
singlet hadrons which are bound states of a single $u$ or $d$ quark and
some isoscalar combination of quarks or antiquarks denoted by $X$
with electric charge $Q_X$ and spin $S_X \not= 0$ to obtain the relations

$$
{{\delta m(D) - \delta m(D^*)}\over{M(D^*) - M(D)}}=
{{\delta m(\Xi) - \delta m(\Xi^*)}\over{M(\Xi^*) - M(\Xi)}}=
{{\delta m(\Xi_b) - \delta m(\Xi_b^*)}\over{M(\Xi_b^*) - M(\Xi_b)}}
\eqno(2a)
$$
$$
{{\delta m(K) - \delta m(K^*)}\over{M(K^*) - M(K)}}=
{{\delta m(B) - \delta m(B^*)}\over{M(B^*) - M(B)}}=
{{\delta m(\Xi_c) - \delta m(\Xi_c^*)}\over{M(\Xi_c^*) - M(\Xi_c)}}
\eqno(2b)
$$

Until recently the errors on the experimental values of the charmed meson
masses were too great to allow a significant test of the prediction
(1). With present data the main source of error is seen to
arise from $\Xi^*$ masses, not the $D^*$'s.
Although the errors are still large $\approx 30\%$ the agreement is
significant. This is seen by noting the very large difference between
the ratios (1) and the experimental$\({\PDG}\)$ value
$ - (7 \pm 3)\times 10^-3 $ for the analogous ratio (2b) for kaons.
This difference is expected since the electric charges of the
strange and charmed quarks have opposite signs while the color
couplings are the same.
Thus the relative signs of the electromagnetic and strong
contributions to the hyperfine splittings are opposite in the two cases.

We now examine these relations at a deeper level
beginning with the DGG constituent quark model$\({\DGG}\)$
in which the hyperfine interaction has the color-spin properties of
one-gluon exchange, is inversely proportional to the product of quark
masses and is treated as a first-order perturbation.
The most naive version apparently neglects all dependence of
hadron wave functions on quark masses and assumes that both
hyperfine interactions and quark magnetic moments scale with the
same effective mass parameters.
Spectacular success has been obtained with this naive approach in using
a ratio of roughly 3/2 between effective quark masses of strange and
nonstrange quarks to explain the observed ratio of hyperfine splittings
between strange and nonstrange mesons and between strange and nonstrange
baryons as well as the ratio of $\Lambda$ and nucleon magnetic moments.
\REF{\LIPCQM}{Harry J. Lipkin,  \plb{233}{446}{1989}}
[{\DGG,\LIPCQM}],

$$ 1.5 = -{{\mu_p}\over{3 \mu_{\Lambda}}} =
{{m_s}\over{m_u}} =
6\cdot {{M(\Lambda) - M(N)}\over{M(\Delta) + M(N)}} + 1 = 1.5
\eqno(3a) $$
$$ 1.5 = {{M(\Delta) - M(N)}\over{M(\Sigma^*) - M(\Sigma)}} =
{{m_s}\over{m_u}} =
{{M(\rho) - M(\pi)}\over{M(K^*) - M(K)}} = 1.6
\eqno(3b) $$
However, the experimental hyperfine splittings in the charmed $D$ mesons
do not show this factor 3/2. Experiment gives
$$ {{m_s}\over{m_u}} = {{M(D^*) - M(D)}\over{M(D_s^*) - M(D_s)}} =
1.01  \not= 1.5  \eqno(3c)  $$

Some insight into this paradox is obtained by noting that
increasing a quark mass not only decreases the strength of the hyperfine
interaction but also increases the value of the wave function at the
origin and therefore increases the matrix element of the interaction.
The two effects are in opposite directions and which is dominant
is not clear a priori. The result (3c) suggests that the two effects
may cancel in the charmed case. But the problem remains why ignoring wave
function effects gives such good results in the lighter quark sector.
Furthermore this model has as yet no rigorous justification from QCD and
the exact meaning of constituent quarks and constituent quark masses
remain unclear.

Although detailed models can be constructed to resolve the paradox (3),
we consider here only the general derivations and implications of the
relations (2) which depend on the very small $u-d$ mass difference.
These relations can be reasonably derived by first-order perturbation
theory and should provide simpler tests of the underlying dynamical
assumptions than other relations connecting states with larger mass
differences.
We investigate various wave function effects neglected in the simple
derivation and note that special problems arise
because there are two independent perturbations. Both the
hyperfine interaction and the mass difference are assumed to be small.
First-order perturbation theory in a single perturbation
automatically takes into account first-order changes in wave functions,
and there is danger of double counting if one also includes wave
function effects in addition to the normal perturbation results. But
with two independent perturbations
it is necessary to specify which first order wave function effects
are automatically included and which are not.

We first obtain some systematic features of the
isospin dependence of hyperfine splittings expressible as inequalities
confirmed by the systematics of the experimental data.
These follow from the simple assumption
that the effect on the wave function is not strong enough to change the
sign of the hyperfine splitting.
We then show that
in a simple constituent quark model the relation (1) is
independent of the structure of the diquark.
Finally we examine the wave function effects in a consistent
double-perturbation
framework, discuss their implications for the relations (2), and
show that their contribution to (1) is still small, even
though they may significantly affect relations like (3).

Using language borrowed from heavy quark symmetry
\REF{\WISGUR}{Nathan Isgur and Mark B. Wise, Phys. Lett. B 232 (1989)
113; B237 (1990) 527.}[{\WISGUR}], let $(Xq)^*$ and $(Xq)$ denote the
two states with ``parallel" and ``antiparallel" spins
of a light quark $u$ or $d$ denoted by $q$ and ``isoscalar brown muck"
denoted by $X$. The two states respectively have
total angular momentum $S_X + 1/2$ and
$S_X - 1/2$. The mass splitting between such a pair of states is assumed
in the DGG model$\({\DGG}\)$
to be due to a hyperfine interaction which has contributions both
from the strong color interaction and from the electromagnetic
interaction. We weaken the DGG assumptions by not necessarily
assuming a constituent quark structure for $X$ nor
that the strong hyperfine interaction is inversely proportional to quark
masses. We only assume that the strong hyperfine interaction
is greater in the $(Xu)$
configuration than in the $(Xd)$ configuration because $m_u < m_d$ and
similarly for the $(Xu)^*$ and $(Xd)^*$ configurations.

It is convenient to define the generalized expression for the ratios
appearing in eq. (1)
$$ F(Q_X) \equiv
{{ \delta m(Xq) - \delta m[(Xq)^*] }\over{M[(Xq)^*] - M(Xq)}}
\eqno(4)
$$
The strong hyperfine interaction between $X$ and $q$ in the ground state
$(Xq)$ configuration is always attractive. It lowers $M(Xq)$
and lowers $M(Xu)$ more than it lowers $M(Xd)$. Thus the
contribution of the strong hyperfine interaction to the ground state isospin
mass splitting has $the$ $same$ $sign$ as the mass difference contribution.
In the excited $(Xq)^*$ configuration the strong hyperfine interaction
between $X$ and $q$ is always repulsive. It raises $M[(Xq)^*]$ and
raises $M[(Xu)^*]$ more than it raises $M[(Xd)^*]$. Thus the
contribution of the strong hyperfine interaction to the excited state
isospin
mass splitting has $the$ $opposite$ $sign$ to the mass difference
contribution.

Thus
$$
\delta_s m(Xq) - \delta_s m[(Xq)^*] \geq 0
\eqno(5a)
$$
where $\delta_s m$ denotes the strong contribution to the isospin
splitting.

For $Q_X <  0$ and in particular for the case $Q_X = -2/3$ the coulomb
and color electric interactions in the (Xu) configuration are both
attractive; i.e. they both have the same sign. Therefore the
corresponding magnetic interactions have the same sign.
In the (Xd) configuration the signs of all electromagnetic
interactions are opposite to those in (Xu).
Thus the electromagnetic hyperfine contributions lower $M(Xu)$ and $M[(Xd)^*]$
while raising $M(Xd)$ and $M[(Xu)^*]$. Their contributions to the isospin
splitting thus also satisfy the relation (5a).
For $Q_X >  0$ and in particular for the case $Q_X = +1/3$ the
electromagnetic contributions have the opposite sign from the case
$Q_X = -2/3$. Their contributions therefore satisfy an inequality opposite to
(5a). This can be expressed for the general case as
$$
-Q_X\{\delta_e m(Xq) - \delta_e m[(Xq)^*]\} \geq 0
\eqno(5b)
$$
where $\delta_e m$ denotes the electromagnetic
hyperfine contribution to the isospin splitting.
We therefore find that for the case $Q_X < 0$ the total isospin
splittings satisfy the inequality
$$
\delta m(Xq) - \delta m[(Xq)^*] \geq 0 \, \, \, {\rm For} \, \, Q_X < 0
\eqno(6a)
$$
while the ratios $ F(Q_X) $ defined by eq. (4) satisfy the inequality
$$ 0 \leq F(-2/3) \geq F(+1/3)
\eqno(6b)
$$
These inequalities are satisfied by experiment.

We now derive the equalities (2) by introducing the following
conventionally accepted additional assumptions;

1. Both the strong and electromagnetic contributions to the hyperfine
splittings are given by the expectation value of the $same$ operator,
not necessarily the value of a wave function at the origin. They
differ only in the values of the coefficients of this operator.

2. The strong hyperfine interaction operator factorizes
into a factor depending only on light quark masses and independent of $X$ and
a factor depending upon the properties of $X$ and independent of light quark
masses. This factorization is found in the DGG model $\({\DGG}\)$.
The quark masses used may be constituent quark masses, current
quark masses or some kind of effective quark masses, all of which
satisfy the inequality $m_u < m_d$.

We consider a ``light quark symmetry" for states described as a single
light quark moving through ``heavy brown muck" containing no valence
light quarks. We assume that the spin and
isospin splittings can be treated by perturbation theory and therefore are
expressible as expectation values of operators describing the
perturbation all with the same unknown brown muck wave function. We use
the Feynman-Hellmann theorem to express the first-order dependences of
the hyperfine splittings on the mass and the charge of the light quark as
expectation values of the derivatives of an unknown operator with respect
to the mass and charge of the light quark. We then find that unknown
``brown muck" properties tend to cancel in the ratio of the isospin
difference between hyperfine splittings to the total hyperfine splitting.
The result is a universal mass formula for this ratio relating all
hadrons in which the ``brown muck" has the same electric charge.

Under the above assumptions the hyperfine mass splittings can be written
$$ \Delta_{hyp} M(Xq) \equiv
M[(Xq)^*] - M[(Xq)] = [S(m_q) - Q_X Q_q E(m_q)]\cdot \bra{Xq} V(X)
\ket{Xq}  \eqno(7)  $$
where $q$ denotes either $u$ or $d$,
$S(m_q)$ and $E(m_q)$ are constants depending on the value of the light quark
mass $m_q$ and independent of $X$, $Q_q$ denotes the charge of the light quark
$q$ and the operator $V(X)$ denotes the hyperfine interaction which depends
upon the properties of X but has
its dependence on the flavor of the light quark $q$ factored out and included
in the coefficients. The constants $S(m_q)$ and $E(m_q)$ and the operator $V$
are defined to make $S(m_q)$, $E(m_q)$ and $\bra{Xq} V(X) \ket{Xq}$ all
positive. This leads to the known positive value of the hyperfine splitting
$M[(Xq)^*] - M[(Xq)]$ and is consistent with the observation that the color and
electromagnetic couplings between $X$ and $q$ have the same sign, attractive,
when $Q_XQ_q < 0$.
Since the electromagnetic interaction is much weaker
than the strong interaction; i.e. $\alpha << \alpha_{strong}$,
$$ E(m_q) << S(m_q) \eqno(8) $$
Then to first order in the small quantities $m_d-m_u$ and
$E(m_q)/S(m_q)$,
the Feynman-Hellmann theorem gives
$$ \Delta_{hyp} M(Xd) - \Delta_{hyp} M(Xu)
= $$  $$
= \left[\left(
{{\partial S(m_q) }\over{\partial m_q}} + \xi S(m_q)\right)
\cdot (m_d-m_u) + Q_X E(m_q) \right] \cdot \bra{Xq} V(X) \ket{Xq}
\eqno(9a)  $$
where the parameter
$$
\xi =
{{\partial }\over{\partial m_q}}\cdot \log \bra{Xq} V(X) \ket{Xq}
\eqno(9b)  $$
expresses the $m_q$ dependence of $\bra{Xq} V(X) \ket{Xq}$. The need for
this correction is discussed and justified below.
Thus
$$ {{\Delta_{hyp} M(Xd) - \Delta_{hyp} M(Xu)}\over
{\Delta_{hyp} M(Xd) + \Delta_{hyp} M(Xu)}}= {1\over {2S(m_q) }}\cdot
\left[\left(
{{\partial S(m_q) }\over{\partial m_q}} + \xi S(m_q)\right)
\cdot (m_d-m_u) + Q_X E(m_q) \right]
\eqno(9c)  $$
When the wave function correction $\xi$ is neglected
the right hand side of eq. (9c) is seen to depend upon the constituent $X$
only via the electric charge $Q_X$ and therefore has the same value for all
constituents $X$ having the same electric charge.

This result (9)
can be expressed in a similar form with electromagnetic mass differences
rather than hyperfine splittings.
This gives the function $F(Q_X)$ defined by eq(4)
as a universal function for all isospin differences between
hyperfine splittings depending only on $Q_X$.
$$ F(Q_X) = -{1\over {S(m_q) }}\cdot \left(
{{\partial S(m_q) }\over{\partial m_q}}\cdot (m_d-m_u) + Q_X E(m_q) \right)
- \xi \cdot (m_d-m_u)
\eqno(10)  $$
When the correction term $\xi$ is neglected
eq. (1) is obtained for the case $Q_X = -2/3$ with $X=\bar c$ and
$X=ss$. We can also include $X=bs$ and generalize eq. (1) to (2a).
Eq. (2b) is obtained
for the case $Q_X= +1/3$ with $X=\bar s$, $X=\bar b$ and $X=cs$.

We now examine the validity of the assumption used in obtaining the
expression (7) for the hyperfine splittings which neglected any effect
of the structure of the brown muck wave function $X$. We consider a
constituent quark model in which $X$ consists of two constituent heavy
quarks denoted by $x_1$ and $x_2$. Then we can write
$$
\Delta_{hyp} M(x_1x_2q) \equiv M[(x_1x_2q)^*] - M[(x_1x_2q)] =
\Delta_{hyp}^s M(x_1x_2q) + \Delta_{hyp}^e M(x_1x_2q)
\eqno(11)  $$
where
$$ \Delta_{hyp}^s M(x_1x_2q) = - s(m_q) \sum_{\alpha=1}^8
\bra{x_1x_2q} g_{\alpha}(q)
[ g_{\alpha}(1)v(x_1) + g_{\alpha}(2)v(x_2) ] \ket{x_1x_2q}
\eqno(12a)  $$
$$ \Delta_{hyp}^e M(x_1x_2q)
= - Q_q E(m_q)\cdot
\bra{x_1x_2q} Q_1 v(x_1) + Q_2 v(x_2) \ket{x_1x_2q}
\eqno(12b)  $$
$E(m_q)$ is the same as in eq. (7) and
$s(m_q)$ is related to $S(m_q)$ by a color factor evaluated below.
Both are constants depending on the value of the light quark
mass $m_q$ and independent of
$x_1$ and $x_2$. The operators $v(x_1)$ and $v(x_2)$
denote the hyperfine interaction at the constituent quark level
which again has its dependence on the flavor of the light quark $q$
factored out and included
in the coefficients. The operators
$g_{\alpha}(q)$, $g_{\alpha}(1)$ and $g_{\alpha}(2)$
denote the eight generators of $SU(3)_{color}$ acting on the particles
$q$, $x_1$ and $x_2$ respectively. Eq. (12a) can be simplified by using
the $SU(3)_{color}$
identity valid for any three-quark color singlet state,
$$ \bra{x_1x_2q} g_{\alpha}(q) + g_{\alpha}(1) + g_{\alpha}(2)
\ket{x_1x_2q} = 0
\eqno(13a)  $$
and noting that
$$C_3(q) = \sum_{\alpha=1}^8 [ g_{\alpha}(q) ] ^2
= \sum_{\alpha=1}^8 [ g_{\alpha}(1) ] ^2
= \sum_{\alpha=1}^8 [ g_{\alpha}(2) ] ^2
\eqno(13b)  $$
is the Casimir operator of color SU(3) for a single quark or antiquark
state. Then
$$ \Delta_{hyp}^s M(x_1x_2q)
= - (1/2) s(m_q) \sum_{\alpha=1}^8
\bra{x_1x_2q} g_{\alpha}(q) [ g_{\alpha}(1) + g_{\alpha}(2) ]
\cdot [ v(x_1) + v(x_2) ] \ket{x_1x_2q} + \Delta_s
= $$
$$ =  (1/2) s(m_q) C_3(q)  \bra{x_1x_2q} v(x_1) + v(x_2) \ket{x_1x_2q}
\eqno(14a)  $$
where
$$ \Delta_s \equiv
- (1/2) s(m_q) \sum_{\alpha=1}^8
\bra{x_1x_2q} g_{\alpha}(q) [ g_{\alpha}(1) - g_{\alpha}(2) ]
\cdot [ v(x_1) - v(x_2) ] \ket{x_1x_2q} = 0
\eqno(14b)  $$
$$ \Delta_{hyp}^e M(x_1x_2q)
= - (1/2) Q_q E(m_q)\cdot [
(Q_1 + Q_2) \bra{x_1x_2q} v(x_1) + v(x_2) \ket{x_1x_2q} + \Delta_e
\eqno(15a)  $$
where
$$ \Delta_e \equiv
- (1/2) Q_q E(m_q) (Q_1 - Q_2)\bra{x_1x_2q} v(x_1) - v(x_2) \ket{x_1x_2q}
\eqno(15b)  $$
Thus
$$ \Delta_{hyp} M(x_1x_2q) =
[S(m_q) - Q_X Q_q E(m_q)]\cdot \bra{x_1x_2q} V(X) \ket{x_1x_2q}
+ \Delta_e
\eqno(16)  $$
where
$$ [S(m_q) =  s(m_q) C_3(q) \eqno(17a)  $$
$$ V(X) = (1/2)[ v(x_1) + v(x_2)] \eqno(17b)  $$
The hyperfine mass splittings for the three-constituent-quark system thus
has the form of the ``brown-muck" formulation (7) except for a
correction term $\Delta_e$ which vanishes identically for the case
$Q_1 = Q_2$ relevant to the expression (1) and can be expected to be
small in the general case since the operator $v(x_1) - v(x_2)$ is
antisymmetric in the particles $1$ and $2$ and has a vanishing
expectation value for a symmetric wave function.

We can now also see how the relation (7) can break down for models
more complicated than simple constituent quark models, where the
internal structure of the ``brown muck"
state $X$ effects the strong and electromagnetic couplings differently.
The expressions (11-17) can  be generalized to the case where
the ``brown muck" state X contains an arbitrary number of quarks and
antiquarks and the result written in the from (16) with parameters
$\Delta_e$ and $\Delta_s$ that vanish in some symmetry limit. However
the symmetry limit is no longer relevant.
There are quarks and antiquarks
having different electric charges, and no simple
permutation symmetry for interchanges of quarks and antiquarks.

We now show that the effect on the relation (1) of the correction term
$\xi$ in (10) is small in the approximation where
$\bra{Xq} V(X) \ket{Xq}$ depends upon $m_q$ only via the reduced mass of
the (Xq) system. This approximation is exact in a two-body
nonrelativistic quark model and can be expected to be reasonably good in
general. For simplicity we assume a power law dependence on the reduced
mass. In this case we can write
$$
\xi = -n {{\partial}\over{\partial m_q}}\cdot \log
\left({1\over{m_q}}+ {1\over{m_X}}\right)
= {n\over{m_q}}\cdot \left(1 - {{m_q} \over{m_q + m_X}}\right)
\eqno(18a)  $$
Then in the DGG model$\({\DGG}\)$
$$\xi + {1\over {S(m_q) }}\cdot \left(
{{\partial S(m_q) }\over{\partial m_q}}\right)= \xi -{1\over{m_q}}
= {{n-1}\over{m_q}} - {{n} \over{m_q + m_X}}
\eqno(18b)
$$
The correction term (18a) is seen to be comparable in magnitude to the
dominant term, but depending strongly on the
ratio $ {{m_q} \over{m_q + m_X}}$. When this ratio is unity; i.e.
$m_X << m_q$ the correction term vanishes and all the good results of
eq. (3) are obtained. When the ratio is small; i.e.
$m_X >> m_q$ the correction term tends to cancel the direct term and
explain the failure of (3) for the charmed sector. We do not attempt
here to carry this argument beyond hand waving. We consider only
the relations (2), where the main correction is
seen to be independent of $m_X$. We now show that
the remaining correction is small. Since
the expression $m_q +m_X$ is approximately equal to the mass of the
hadron $(Xq)$, we can express the difference in the value of the function
$F(Q_X)$  (4) for two hadron states denoted by $A$ and $B$ having the
same value of $Q_X$ as
$$ F(Q_X)_B - F(Q_X)_A  \approx
{{n (M_A - M_B)} \over{M_A  M_B}}
\cdot (m_d-m_u)
\eqno(19)  $$
For the case of the $D$ and $\Xi$ hadrons of eq. (1) the correction
to the strong contribution is $\approx 8\%$ if we take n=1 and
$m_q =360$ MeV and the correction to (1) is $\approx 4\%$ if the strong
and electromagnetic contributions are roughly equal.
The effect can be considerably larger in the comparison
of the $K$ and $B$ systems. It will be interesting to check this
experimentally.

We now return to the validity of the correction term $\xi$ in eqs.
(9-10) expressing the $m_q$ dependence of the matrix element
$\bra{Xq} V(X) \ket{Xq}$. The Feyman-Hellmann theorem tells us to ignore
wave function effects in first-order perturbation theory because all
such effects cancel as a result of the variational principle. Therefore
the first-order change in $\bra{Xq} V(X) \ket{Xq}$ must be canceled
by another contribution. This other contribution arises from the very
small spin-dependent change in the wave function produced by a change in
$m_q$ because the wave function is an eigenfunction of the total
Hamiltonian which includes the hyperfine interaction. This change in the
wave function produces a $spin-dependent$ energy change via the
$spin-independent$ part of the Hamiltonian; e.g. the kinetic energy. The
Feynman-Hellmann theorem and the variational principle tell us that
effects of this kind must exactly cancel the contribution to the energy
from the change in $\bra{Xq} V(X) \ket{Xq}$ to first order. This
argument suggests that the correction term should not be introduced in
eqs. (9-10).

However, there are
really four independent matrix elements and two perturbations: the
mass difference and the hyperfine interaction. We
are calculating a second order cross term in a double
perturbation series. The Feynman-Hellmann theorem cannot be used twice
to justify using the same zero-order wave function in the same zero-order
Hamiltonian for four matrix elements.

We now clarify this point explicitly and define a consistent
perturbation calculation for the second order cross term.
Consider a general hamiltonian having the form
$$ H = T(m_q) + V  + {{\epsilon}\over{m_q}} V_{hyp}     \eqno(20) $$
where $T(m_q)$ is the kinetic energy of the light quark but can be
generalized to include other contributions which depend
upon $m_q$, $V$ denotes all contributions to the Hamiltonian which do
not depend explicitly on $m_q$, $\epsilon$ is a small parameter
specifying the strength of the hyperfine interaction, all the explicit
dependence of the hyperfine interaction on $m_q$ is in the factor
${1\over{m_q}}$, and the operator $V_{hyp}$ contains no explicit
dependence on $m_q$. We are
interested in the dependence of $\langle H \rangle$ on $m_q$ and
$\epsilon$ to first order in the product $\epsilon \cdot \delta m_q$,

Using the Feynman-Hellmann theorem in two ways gives
$$ {{\partial \langle H \rangle }\over{\partial \epsilon }}
= {{\langle V_{hyp} \rangle}\over{m_q}}
   \eqno(21a)  $$
$$ m_q {{\partial \langle H \rangle }\over{\partial m_q}}
=  m_q \left\langle {{\partial T }\over{\partial m_q}}\right\rangle
- {{\epsilon}\over{m_q}} \langle V_{hyp} \rangle    \eqno(21b)  $$
where the variational principle as expressed by the Feynman-Hellmann
theorem allows us not to consider the derivatives of the matrix elements.
We now obtain the crossed second derivative in two ways by
direct differentiation of eqs. (21).
Here there is no variational principle and the derivatives of the matrix
elements must also be considered.
$$ m_q {{\partial^2 \langle H \rangle }\over{
\partial m_q \partial \epsilon }}
= - {{\langle V_{hyp} \rangle }\over{m_q}} +
 {{\partial}\over{\partial m_q }} \cdot \langle V_{hyp} \rangle
\eqno(22a)  $$
$$ m_q {{\partial^2 \langle H \rangle }\over{
\partial \epsilon \partial m_q }}  =
- {{\langle V_{hyp} \rangle }\over{m_q}} +
m_q
{{\partial }\over{\partial \epsilon }}\cdot
\left\langle {{\partial T }\over{\partial m_q}}\right\rangle
\eqno(22b)  $$
These results (22a) and (22b) both give the same dominant term
$- {{\langle V_{hyp} \rangle }\over{m_q}} $ used in eqs. (9) when
wave function effects are neglected.
The result (22a) is just the full eq. (9) and justifies the use of
the correction term (9b) for the
change in the matrix element of the hyperfine interaction. Combining
eqs. (22) gives
$$  {{\partial}\over{\partial m_q }} \cdot \langle V_{hyp} \rangle
= m_q  {{\partial }\over{\partial \epsilon }}\cdot
\left\langle {{\partial T }\over{\partial m_q}}\right\rangle
\eqno(23)  $$
This condition relating the mass dependence of the hyperfine matrix
element and the effect of the hyperfine interaction on the kinetic energy
seems at first very peculiar. But it must hold for any model described by
the Hamiltonian (20) since it follows from calculating the same second
derivative in two ways. It may be useful as a consistency check on
detailed model calculations.

We now show how the variational principle relates kinetic and
potential energies in the manner required by eq. (23) in
the case of a nonrelativistic quark model and a short-range hyperfine
interaction which can be written
$$ \langle V_{hyp} \rangle
= \int_o^{r_o} r^2 dr |\psi(r)|^2 v(r)
\approx |\psi(0)|^2 \int_o^{r_o} r^2 dr v(r)
\eqno(24a)  $$
$$
\left\langle r {{dV_{hyp}}\over{dr}} \right\rangle \approx
|\psi(0)|^2 \int_o^{r_o} r^3 dr {{dv}\over{dr}} =
- |\psi(0)|^2 \int_o^{r_o} 3 r^2 dr v(r) =
- 3 \langle V_{hyp}
\rangle
\eqno(24b)  $$
where $\psi(0)$ denotes the wave function at the origin and v(r) is a
short range potential which vanishes for $r \geq r_o$ and we neglect
the variation in the wave function for $r \leq r_o$.
The virial theorem gives
$$ m_q \left\langle {{\partial T }\over{\partial m_q}}\right\rangle
= - {{\mu}\over{m_q}} \cdot \langle T_{rel} \rangle =
- {{\mu}\over{2 m_q}} \cdot \left\langle r \cdot {{dV}\over{dr}}
+ {{\epsilon}\over{m_q}} \cdot r {{dV_{hyp}}\over{dr}} \right\rangle
\eqno(25)  $$
where $T_{rel}$ denotes the kinetic energy and
$\mu$ the reduced mass of the relative motion in the (Xq) system.
Substituting eqs. (24-25) into the condition (23) gives
$$ {{\langle V_{hyp} \rangle}\over{|\psi(0)|^2}}
\cdot {{\partial (|\psi(0)|^2)} \over{\partial m_q }} \approx
{{3 \mu}\over{2 m_q^2}} \cdot \langle V_{hyp} \rangle
- {{\mu}\over{2 m_q}} \cdot
{{\partial }\over{\partial \epsilon }}\cdot
\left\langle r \cdot {{dV}\over{dr}}
\right\rangle
\eqno(26)  $$
This condition must be satisfied in any nonrelativistic quark model with
a short range hyperfine interaction and
is seen by inspection to be satisfied for the case of a
logarithmic potential model where $ r \cdot {{dV}\over{dr}} $
is a c-number independent of $\epsilon$ and
$|\psi(0)|^2$ varies as $\mu^{(3/2)}$.

We thus conclude that the relation (1) is valid under very general
assumptions going beyond the naive DGG model and that it is of interest
to improve the precision of the mass measurements. The remaining
relations (2) of this type are more sensitive to wave function effects
but should still be approximately valid. The way in which wave function
effects depending upon reduced mass can explain the paradox (3) is
qualitatively indicated in eq. (18a). But a more quantitative
explanation requires a more detailed model and probably a
relativistic calculation.

Stimulating and clarifying discussions with Nathan Isgur and Jonathan L.
Rosner and Sheldon Stone are gratefully acknowledged. This research was
partially supported by the Basic Research Foundation administered by the
Israel Academy of Sciences and Humanities and by grant No. 90-00342 from
the United States-Israel Binational Science Foundation (BSF), Jerusalem,
Israel.

\refout
\end